\documentclass{PoS}
\usepackage{soul}
\usepackage{url}
\usepackage{ae,aecompl}
\usepackage{cleveref}
\usepackage{comment}
\usepackage{pdflscape}
\usepackage[square,numbers,sort&compress]{natbib}

\usepackage{array}

\usepackage{graphicx}	
\usepackage{amsmath}	
\usepackage{amssymb}	
\usepackage{subfig}
\usepackage{caption}

\usepackage{floatrow}
\usepackage{hvfloat}
\title{Extreme high-energy peaked BL Lac objects and their TeV gamma-ray emission: are they a homogeneous population?}

\ShortTitle{Unveiling the TeV gamma-ray nature of EHBLs}

\author{\speaker{L. Foffano}\\
University of Padova  (Physics and Astronomy department) and Istituto Nazionale di Fisica Nucleare (INFN), via Marzolo 8, 35131 Padova, Italy\\
E-mail: \email{luca.foffano@phd.unipd.it}}

\author{E. Prandini\\
Istituto Nazionale di Astrofisica (INAF), vicolo dell'osservatorio 3, 35131 Padova, Italy and Istituto Nazionale di Fisica Nucleare (INFN), Italy}

\author{A. Franceschini\\
University of Padova  (Physics and Astronomy department), via Marzolo 8, 35131 Padova, Italy\\
Istituto Nazionale di Astrofisica (INAF), vicolo dell'osservatorio 3, 35131 Padova, Italy
}

\author{S. Paiano\\
Istituto Nazionale di Astrofisica (INAF), vicolo dell'osservatorio 3, 35131 Padova, Italy\\
} 

\abstract{Extreme high-energy peaked BL Lac objects (EHBLs) are an emerging class of blazars with exceptional spectral properties. In blazars,  the spectral energy distribution (SED) is dominated by the non-thermal emission of the relativistic jet, and consists of two main broad humps. For the EHBLs, these two components peak in the X-ray and GeV-TeV bands, respectively. Although the number of TeV detected extreme blazars is very limited, recent observations by Imaging Atmospheric Cherenkov Telescopes (IACTs) have revealed that in some of them the energy of the second peak exceeds several TeV (e.g. 1ES 0229+200). Their exceptional hard TeV spectra represent a challenge for the standard leptonic modeling, and a possible hadronic contribution may make these objects high-energy neutrinos producers. Moreover, they are important for the implications on the indirect measurements of the extragalactic background light and of the intergalactic magnetic field. 
In this contribution, we perform a comparative study of the multi-wavelength spectral energy distributions of a sample of hard X-ray selected EHBL objects. The analysis suggests that the EHBL class is not homogeneous, and a possible sub-classification may be unveiled with TeV gamma-ray observations of the candidates.
With the purpose of increasing their number and settle their statistics, we discuss the potential detectability of the currently undetected TeV-emitting EHBLs in our sample by current and next generation of  IACTs.
}

\FullConference{36th International Cosmic Ray Conference -ICRC2019-\\
		July 24th - August 1st, 2019\\
		Madison, WI, U.S.A.}

\begin{document}


\section{Introduction}
\noindent
Blazars are active galactic nuclei (AGN) emitting enhanced non-thermal radiation covering the entire electromagnetic spectrum from radio up to gamma rays.
This emission is mainly due to synchrotron processes involving the electrons accelerated within the relativistic jet, and results in the first low-energy hump of the spectral energy distribution (SED) of the blazar. In the simple Synchrotron Self-Compton model (SSC, e.g. \cite{Maraschi:SSC,Tavecchio:SSC}), such synchrotron radiation is scattered up to the highest energies via Inverse Compton interaction \citep{IC} by the very same population of accelerated electrons, producing the second high-energy hump of the SED. Alternative models interpret this high-energy hump as due to the Inverse Compton scattering of ambient photon fields coming from nearby regions (External Compton model, \cite{EC}), or as due to different combinations of leptonic and hadronic processes taking place in the blazar jet and in the surrounding environment.

In the blazar family, the BL Lac objects are characterized by absence of significant emission and absorption lines in the optical spectrum. They are generally classified depending on the frequency of the energy peak of the synchrotron photons emitted in their relativistic jets.
The objects in this class that are able to produce the highest photon energies are called extremely high-energy peaked BL Lac objects (EHBLs, \cite{Costamante:2001pu}), and are defined by a synchrotron peak located at energies above about~0.3~keV~($\simeq 10^{17}$~Hz) \citep{costamante2002, fermi-bright-blazars}.

\begin{figure}
\centering
\includegraphics[width=\columnwidth]{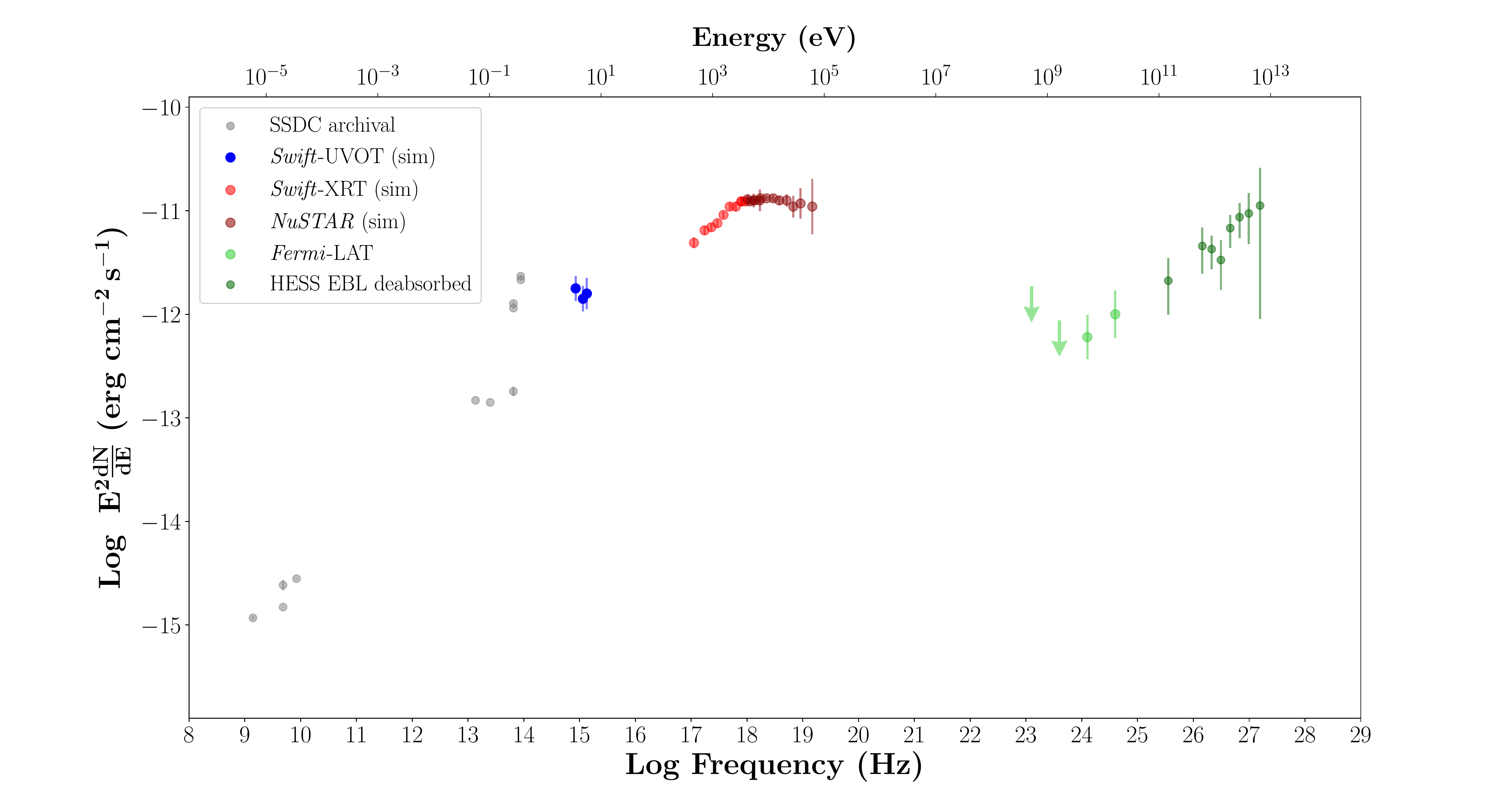}
\caption{Multi-wavelength SED of the archetypal EHBL called 1ES~0229+200. Here we report in grey archival data from SSDC website$^1$.
From \cite{Nustar_EHBLs}, we show in light red the data points from \mbox{\emph{Swift}-XRT}, in dark red the data points from \emph{NuSTAR}, in light green the updated analysis of \mbox{ \emph{Fermi}-LAT} data over about ten years of observation. Finally, in dark green we report the VHE data from~\cite{0229_hess_points}. The plotted data are already corrected for EBL absorption with the model by \cite{Franceschini17} to show the intrinsic spectrum of the source.
}
\label{fig:1ES0229+200}
\end{figure}

The archetypal object of this class is  1ES~0229+200. It has been characterized in detail through  dedicated multi-wavelength (MWL) campaigns performed during the last years \citep[e.g.][]{Kaufmann:2011-1es0229}.
Its \mbox{broad-band} SED, reported in Figure~\ref{fig:1ES0229+200}, shows some of the most important spectral properties characterizing EHBLs: a synchrotron peak  located at energies about 10 keV \citep{Nustar_EHBLs}, and a high-energy hump exceeding some TeV.

The maximum energy of the high-energy hump reaching the TeV gamma-ray band makes it particularly suitable for the Imaging Atmospheric Cherenkov Telescopes (IACTs). 
The detection of the peak emission at the highest energies and its spectral characterization seem to be crucial in order to disclose the extreme emission mechanism of such objects.

However, in EHBLs the location in the SED at the highest energies of both the synchrotron and the high-energy humps are associated to a lower flux through all the entire broad-band spectrum. For this reason, the detection of these objects in the very-high-energy gamma-ray band (VHE, energies above 100\,GeV)  with the current generation of IACTs is  challenging, and the number of objects that are currently classified as TeV gamma-ray detected EHBLs is poor. 
Further observations of these objects have to be performed on good samples of EHBLs that are also good candidates to be detected in TeV gamma rays. 

In this work from \cite{foffano2018}, we investigate a sample of EHBLs selected from a hard X-ray survey.
We analyze the broad-band spectral properties of the sample including the TeV gamma-ray band, when available, and for the TeV gamma-ray undetected sources we assess the detectability for the current and  next generation of IACTs.

\section{Sample}

\noindent
EHBLs peak in the hard X-ray band with their synchrotron emission. Considering that they are the only class of blazars able to produce synchrotron photons up to that energy band, the search for EHBLs through all-sky hard X-ray surveys represents a promising selection criterion. In this work, we analyzed the \emph{Swift}-BAT 105-months catalog \citep{BATcatalog105}, that represents - up to now - the most sensitive and uniform all-sky survey in the  hard  X-ray band   (14-195 keV), provided by the Burst Alert Telescope (BAT) instrument \citep{BATinstrument} on board of the \emph{Neil Gehrels Swift} satellite \citep{swiftsatellite}. We selected sources classified as ``beamed AGNs'' in this catalog, and then we cut for sources having detection also in the \emph{Fermi}-LAT 3LAC catalog \citep{3LACcatalog}. This way we selected a good sample of blazars with enhanced emission at hard X-rays and emitting also at high-energy (HE, energies between 100 MeV and 100 GeV) gamma rays.


At the end, the  selection procedure provided us a final sample of 32 EHBL objects. After having collected their archival MWL SEDs, we analyzed and compared their broad-band spectral properties. The result was that all sources present substantially compatible spectral properties through their broad-band spectra, except for the TeV gamma-ray band in which there are increasing differences depending on  the energy.

Considering that - within the final sample of 32 objects -  18 sources are already TeV gamma-ray detected, we had the opportunity to analyze their broad-band spectral properties up to the highest available gamma-ray energies. The result from this analysis lead us to study the TeV gamma-ray undetected sources, by performing extrapolations of the TeV gamma-ray undetected sources in order to estimate their detectability at VHE.

\footnotetext[1]{\url{https://tools.ssdc.asi.it/index.jsp}}

\begin{figure}
\centering
\includegraphics[width=\columnwidth]{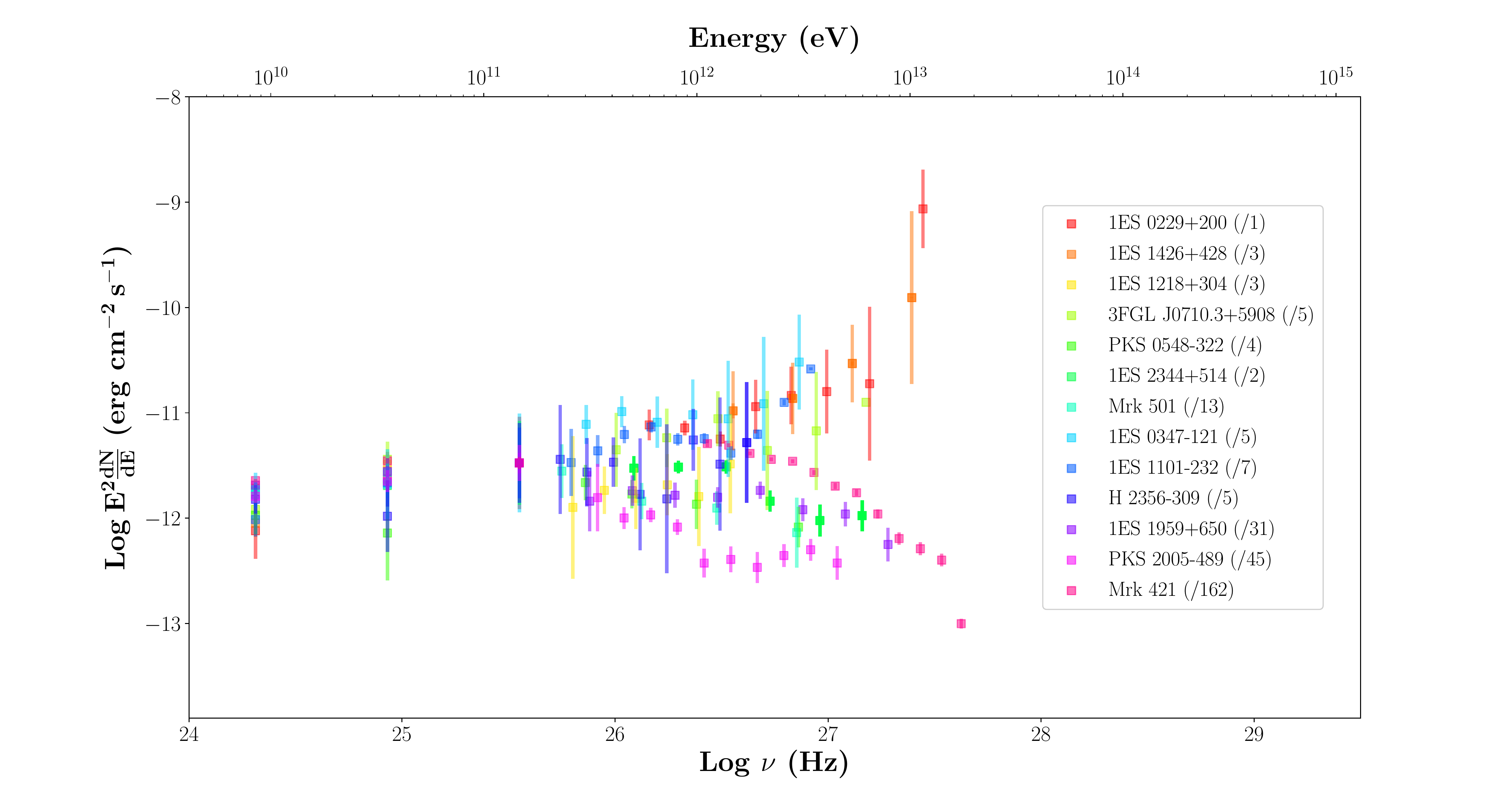}
\caption{Focus on the HE-VHE gamma ray range of the TeV gamma-ray detected sources in \cite{foffano2018} with fluxes normalized to the 1ES~0229+200 flux at 147 GeV with $1.93 \cdot 10^{-12}$ erg $\text{cm}^{-2}$ $\text{s}^{-1}$. In parentheses we report the ratio of the spectra of each source with respect to the normalization chosen in the 1ES~0229+200 spectrum. The plotted data are already corrected for EBL absorption with the model by \cite{Franceschini17} to show the intrinsic spectrum of the source. Figure adapted from \cite{foffano2018}.}
\label{fig:HEspectra}
\end{figure}


\subsection{TeV gamma-ray detected  sources}

\noindent
In the sample of 18 TeV gamma-ray detected sources, their spectral properties in the TeV \mbox{gamma-ray} band show crucial differences. Once normalized the SED at a reference flux level, a focus in the HE gamma-ray spectra of these objects leads to see a substantial compatibility of the HE spectra and an interesting divergence of the corresponding TeV gamma-ray spectra. As reported in \Cref{fig:HEspectra}, some sources present a curved gamma-ray spectrum that peaks at about few hundred of GeV. Conversely, some other sources show a continuously increasing spectrum from HE up to the VHE gamma rays, and in some cases the peak is not even detected.

\begin{figure}
\centering
\includegraphics[width=\columnwidth]{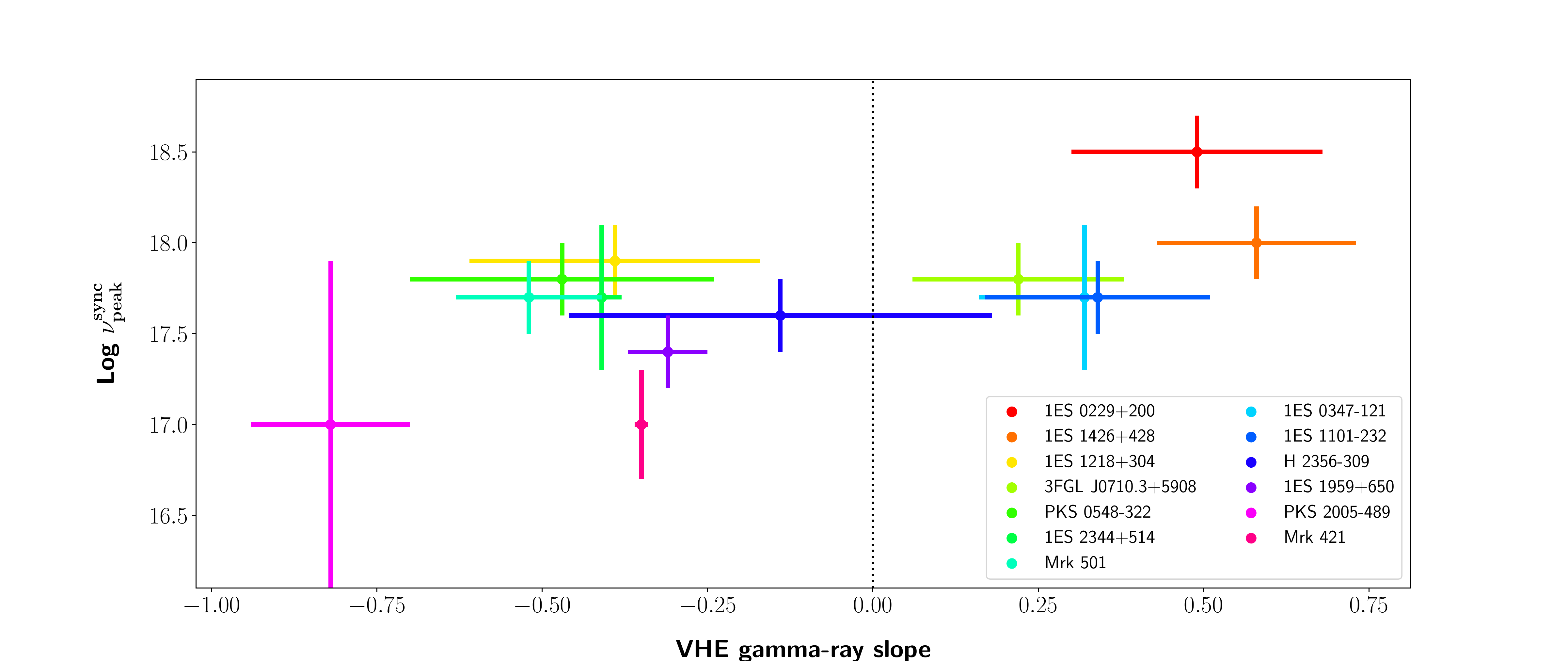}
\caption{Synchrotron peak frequency of the TeV gamma-ray detected sources   with respect to the distribution of their slopes in the 0.1-10 TeV range. Figure adapted from \cite{foffano2018}.}
\label{fig:TeV_slopes_vs_synchropeak_LUM}
\end{figure}

Such behaviour has been analyzed quantitatively by computing the spectral index in different parts of the high-energy hump, and verifying the differences of its value along the HE and VHE parts of the spectrum.
The results - reported in \Cref{fig:TeV_slopes_vs_synchropeak_LUM} - show a clear clustering of sources with continuously increasing gamma-ray spectrum up to the VHE, while some other sources present clearly the detection of their high-energy hump peak at few hundreds of GeV. While in the first group all the already known ``hard-TeV'' blazars  are present (sources characterized by flux stability, no evidence of flaring episodes, and high-energy peak located at energies about 10 TeV or above, see e.g. \cite{Nustar_EHBLs}), in the second group we have sources that show moderate flux variability and flaring episodes during which they show EHBL behaviour raising their synchrotron peak above 0.3 keV (we call them ``HBL-like EHBLs'').

Between these two classes, there might be some sources that are peaking at an intermediate energy at about few TeV, like H~2356-309. Such sources may represent the gradually different location of the high-energy peak in the EHBL class. However, we need more statistics in order to confirm this scenario.

\subsection{TeV gamma-ray undetected  sources}

\noindent
In the final sample of objects, 18 sources are not detected yet in the TeV gamma-ray band. For this reason, we analyzed their HE gamma-ray spectra in order to understand if they might present spectral properties that at VHE are compatible with one of the two previous groups in the TeV gamma-ray detected sample. Additionally, this leads us to check if they are eventually detectable at TeV gamma-ray energies with the current and the next generation of Cherenkov telescopes. 

For this reason, we extrapolated the newly characterized ten-years \emph{Fermi}-LAT spectrum by assuming a power-law function with exponential cut-off. 
The choice of this function is driven by the observational evidence that the EHBLs  we are looking for show power-law spectra up to the deep TeV gamma-ray range \cite{Nustar_EHBLs}, with the possible presence of an high-energy peak that we describe with the exponential cut-off.

In this analysis, a particular attention has been focused on the redshift of these sources. In fact, the EBL interaction strongly suppresses the TeV gamma-ray flux of these objects, and makes them substantially undetectable by the Cherenkov telescopes when the redshift is too high (e.g.~above~0.5, for reasonable amount of integration time of the order of 50~h).
The extrapolations were thus absorbed by EBL interaction with the known values of redshift for each source with the model by \cite{Franceschini17}. 

The resulting set of good candidates to be observed and characterized with  the IACTs is composed by sources like PKS~0352-686, 1RXS~J225146.9-320614, BZB~J1417+2543, and BZB~J0244-5819. They represent good targets also for the MAGIC, H.E.S.S., and VERITAS telescopes, likely detectable in less than 50~h of good quality observations.


\section{Conclusions}

\noindent
In this work, we present a new set of 32 EHBLs selected through the analysis of hard X-ray and HE gamma-ray surveys. 
The analysis of their MWL SEDs reveals that the EHBLs have generally compatible  properties through their broad-band spectra, except for the TeV gamma-ray band.
For this reason, we splitted the sample in the 18 TeV gamma-ray detected sources and 14 TeV \mbox{gamma-ray} undetected sources. The analysis of the first group lead to an increasing divergence of the gamma-ray spectra when reaching TeV energies. Such spectral differences might be indicating a sub-classification of the EHBL category.

The extrapolation of the HE gamma-ray spectra of the TeV gamma-ray undetected sample of sources allowed us to assess the detectability of all  sources by  current and next generation of Cherenkov telescopes.



\bibliographystyle{abbrvnat}   
\bibliography{bibpaper}

\end{document}